\documentclass[9pt]{article}

\usepackage[margin=2cm]{geometry}
\usepackage{graphicx}
\usepackage{amssymb}
\usepackage{amsmath}
\usepackage{authblk}

\newcommand\Beq{\begin{eqnarray}}
\newcommand\Eeq{\end{eqnarray}}

\newcommand{\eq}[1]{Eq.~(\ref{#1})}
\newcommand{\eqs}[2]{Eqs.~(\ref{#1})~\&~(\ref{#2})}
\newcommand{\eqss}[2]{Eqs.~(\ref{#1})--(\ref{#2})}

\newcommand{\Eq}[1]{Eq.~(\ref{#1})}

\newcommand{\Fig}[1]{Fig.~\ref{#1}}

\newcommand{\ie}{\textit{i.e.}, }
\newcommand{\etal}{\textit{et al.~}}
\newcommand{\sol}{\odot}
\newcommand{\cz}{\text{c.z.}\!}
\newcommand{\sn}[2]{$#1 \times 10^{#2} $}
\newcommand{\trow}[4]{{\footnotesize #1} & {\footnotesize $#2$} & {\footnotesize #3}  & {\footnotesize #4} \\ }

\newcommand{\pd}[1]{\partial_{#1}}
\renewcommand{\O}[1]{\approx\!\text{#1}}

\newcommand{\Ro}{\text{Ro}}
\newcommand{\RoD}{\Ro_{\text{d.}\!}}
\newcommand{\RoC}{\Ro_{\text{c.}\!}}
\newcommand{\RoB}{\Ro_{\text{b.}\!}}

\newcommand{\f}{2 \Omega }
\newcommand{\z}{\hat{z} }
\renewcommand{\L}{\ell} 
\renewcommand{\H}{\mathrm{\textsl{H}}_{\cz}}
\newcommand{\h}{\mathrm{\textsl{H}}_{\rho}}
\renewcommand{\u}{|u|} 
\renewcommand{\t}{\tau} 

\newcommand{\dd}[1]{\,\text{d}{#1}}
\newcommand{\grad}{\nabla}
\renewcommand{\div}{\nabla \cdot}
\newcommand{\curl}{\nabla \times}

\title{\bf The rotational influence on solar convection}

\author[a]{Geoffrey M. Vasil\footnote{Correspondences: geoffrey.vasil@sydney.edu.au}}
\author[b]{Keith Julien}
\author[b,c]{Nicholas A. Featherstone}

\affil[a]{\footnotesize{University of Sydney School of Mathematics and Statistics, Sydney NSW 2006, Australia}}
\affil[b]{Department of Applied Mathematics, University of Colorado, Boulder, CO 80309-0526, USA}
\affil[c]{Southwest Research Institute, Department of the Space Studies, Boulder, CO 80302, USA}

\begin{document}

\maketitle

\begin{abstract}
This paper considers the dominant dynamical, thermal and rotational balances within the solar convection zone. 
The reasoning is such that:
Coriolis forces balance pressure gradients. 
Background vortex stretching, baroclinic torques and nonlinear advection balance jointly.
Turbulent fluxes convey what part of the solar luminosity that radiative diffusion cannot. 
These four relations determine estimates for the dominant length scales and dynamical amplitudes strictly in terms of known physical quantities.
We predict that the dynamical Rossby number for convection is less than unity below the near-surface shear layer, indicating strong rotational constraint. 
We also predict a characteristic convection length scale of roughly 30 Mm throughout much of the convection zone.
These inferences help explain recent observations that reveal weak flow amplitudes at 100-200 Mm scales. 
\end{abstract}

\section{Introduction}

Turbulent thermal convection dominates the energy transport throughout the Sun's outer envelope. 
In recent years, puzzling disagreements have arisen between observations, models and theory regarding the amplitude and structure of convection \cite{HanasogeEtal2012,GreerEtal2016,HanasogeEtal2020}. 
Some work has begun to address the situation \cite{MieschEtal2012,FeatherstoneHindman2016b}.
However, the discrepancy is not settled and has come to be called the \textit{convective conundrum}\footnote{Mark Rast 2014, private communication}. 

In this paper, we demonstrate how the dominant dynamical balances in the solar interior make first-principle predictions for the spatial scale and amplitude of deep solar convection. 
Our analysis shows that interior flows likely exists in a Quasi-Geostrophic (QG) state, with a joint Coriolis-Inertial-Archimedean balance (CIA; see, e.g., \cite{GilletJones2006,SchwaigerEtal2019,AurnouEtal2020}) remaining after accounting for leading-order geostrophy\footnote{
The geodynamo community has started referring to QG-MAC balance, which includes magnetism. 
The distinction is unnecessary here because magnetic energies in the Sun are not greater than kinetic energy; \ie MAC $\sim$ CIA.}.
Rotation strongly influences solar convection as a result. 
Our results put on a firm theoretical footing the earlier suggestions of Miesch \etal 2012 and Featherstone \& Hindman 2016 \cite{MieschEtal2012,FeatherstoneHindman2016b}. 
We also corroborate and provide context to the observations of Hanasoge \etal 2012, 2020 \cite{HanasogeEtal2012,HanasogeEtal2020}. 
New estimates leave the flow amplitude only somewhat smaller than previous mixing-length models.
Rotational influence in the Sun most prominently affects the dominant flow length scale.

\section{Solar convective processes}

Apart from negligible friction, fluid conserves angular momentum as it traverses the solar envelope.
Angular-momentum transport occurs throughout the interior and generates differential rotation comprising a fast-equator and slow poles \cite{Howe2009,ThompsonEtal2003}.  
Angular-momentum redistribution also drives a large-scale north-south meridional circulation \cite{GastineEtal2013,GastineEtal2014,GuerreroEtal2013,FeatherstoneMiesch2015}.
Long-term observations document the near-surface meridional flow.
However, its depth dependence is much less clear, with different helioseismic techniques yielding different results \cite{HathawayEtal2003,HathawayEtal2010,Ulrich2010,KholikovEtal2014,JackiewiczEtal2015,GizonMC2020}.

Large-scale plasma motions must play a pivotal role in the stellar dynamo process. 
The latitudinal and radial shear provides a poloidal-to-toroidal conversion mechanism;  \ie the $\Omega$-effect \cite{Ossendrijver2003}. 
Meridional circulation modulates the distribution of sunspots and may also establish the cycle timing  \cite{DikpatiMC99,YeatesEtal2008,MuozJaramilloEtal2009}. 
Helical flow generates a mean electro-motive force (\ie the $\alpha$-effect), which provides a toroidal-to-poloidal dynamo feedback \cite{Moffatt1978,OlsonEtal1999,Davidson2001}. 
Any new information concerning interior fluid motions will produce valuable insight into the operation of the Sun's magnetic cycle.

\subsection*{Photospheric Convection}

Several decades of observations have revealed much about solar surface convection \cite{Hart1956,LeightonEtal1962}.
Driven by fast radiative cooling, granulation dominates the radial motion at the solar surface. 
Roughly 1 Mm in horizontal size, granulation produces a strong power-spectrum peak at spherical-harmonic degree $\O{1000}$ in radial Dopplergrams \cite{HathawayEtal2000,HathawayEtal2015}.
Significant power also exists at the $\O{30\,Mm}$ supergranular scale, whose associated motions are mainly horizontal and best observed in limb Dopplergram spectra \cite{LawrenceEtal1999}.

Photospheric power decreases monotonically for scales larger than supergranulation.
From non-rotating intuition \cite{AhlersEtal2009}, we might expect power to peak at $\O{100-200\,Mm}$, rather than the $\O{30\,Mm}$ supergranular scale. 
The results of \cite{HathawayEtal2013,Hathaway2020} indicate that deep-rooted fluid motions do persist on larger scales.  
Against expectations, however, these motions appear weak compared to the smaller-scale supergranular and granular flows.  
A great deal of theory and simulation work has attempted to solve the supergranulation problem.
To date, no model self-consistently demonstrates how the supergranular scale might arise. 
We direct the reader to the recent review by Rincon \etal 2017 \cite{RinconEtal2017} for a thorough discussion of this topic. 
Our focus here is on the apparent lack of large-scale power as expected from non-rotating convection.  
We suggest, as have others \cite{Lord14,FeatherstoneHindman2016b}, that the supergranular scale results from suppression of power on large scales, rather than through preferential driving at that spatial scale.

\subsection*{Sub-photospheric Convection}

Local helioseismic techniques can probe subsurface convection directly (e.g., time-distance \cite{DuvallEtal1993}, ring-diagram analysis \cite{BasuEtal1999}, holography \cite{LindseyBraun1997}).
Historically, these methods have largely been limited to $\O{30\,Mm}$ depth and do not sample flow below the near-surface shear layer. As a result, numerical simulations play a substantial role in describing the dynamical balances in the deep convection zone.  

Initially, nonlinear simulations of the full, rotating solar convection zone seemed to reproduce the Sun's differential rotation profile. 
Those results suggested (as expected from non-rotating intuition) that convective power peaks at $\O{100\,Mm}$ scales with $\O{100\,m/s}$ flow-speed amplitudes \cite{BrunToomre2002,MieschEtal2006}.  
Limitations of those results began to appear, however, with systematic magnetohydrodynamic studies in the ensuing decade. 
These studies found that only systems with $\O{10}\times$ weaker flows or equivalently, those which rotated $\O{10}\times$ faster, were able to produce coherent magnetic fields and periodic magnetic cycles in analogue to the Sun \cite{Ghizaru10,Racine11,Brown11,Kapyla12,Warnecke14,Nelson15}.
Moreover, simulations with more extreme parameters and large-scale power can generate anti-solar differential rotation, with slow equator and fast poles, \cite{GastineEtal2013,Gastine13,Guerrero13,Kapyla14,FeatherstoneMiesch2015,BrunBrowning2017}. 
We also note that some recent observations suggest the Sun may lie near a boundary between these two basins in parameter space \cite{MetcalfeEtal2016}.

 While most local helioseismic analyses focus on the near-surface shear layers, techniques have been developed to probe more deeply-rooted flow structures, such as solar meridional circulation \cite{Hartlep13,ZhaoEtal2013}.  A notable puzzle arose following the deep-focusing time-distance analysis of Hanasoge \etal 2012 \cite{HanasogeEtal2012}.  This work placed a roughly $1\,\text{m/s}$ upper limit on the $\O{100\,Mm}$ scale flow amplitudes at a depth $\O{60\,Mm}$. 
Subsequently, Greer \etal 2016 \cite{GreerEtal2016} sampled deeply enough to compare directly against the time-distance results at a depth of 30\,Mm.  
Rather than a nondetection, this effort yielded measured flows that were 10--100$\times$ larger on those spatial scales.
The disagreement between these results remains unresolved. 

As an alternative to helioseismic measurement, the gyroscopic pumping effect \cite{MieschEtal2012} could, in principle, map the structure of deep convection. 
However, the technique requires accurate measurement of differential rotation and deep meridional circulation.
Unfortunately, the current ambiguity in meridional circulation measurements makes this strategy presently impossible \cite{HathawayEtal2003,HathawayEtal2010,Ulrich2010,KholikovEtal2014,JackiewiczEtal2015,GizonMC2020}.

\subsection*{The Convective Conundrum}

As a summary, we believe the following are all closely related questions
\begin{itemize}
\setlength\itemsep{-0.1cm}
\item{Where does supergranulation come from?}
\item{Why are classic ``giant cells'' not observed?}
\item{Why and exactly how do observations seem to contradict numerical models?}
\end{itemize}
These questions all essentially ask the same thing: ``Where is the large-scale convective power?'' 
Featherstone and Hindman 2016 \cite{FeatherstoneHindman2016b} pointed out that rotational effects provide a natural explanation for all three of these questions.  
Based on rotational effects, they suggested that the horizontal scale of deep convection must be no larger than the supergranular scale. 
They also estimated interior convective speeds significantly weaker than previously predicted.
 
That work was numerical in nature, however.  
It did not capture, nor describe theoretically, the transition between near-surface and deep-seated convection.
Here, we carry out a careful theoretical analysis of rotational effects and demonstrate how, indeed, the influence of rotation provides a striking response to these questions of solar convection.

\section{Analysis}

Our goal is to estimate the dominant forces and their relative magnitudes. 
Our program is to manipulate the equations of motion into a form that exposes the principal dynamical balances as much as possible.  

\subsection*{Background}

In a coordinate frame with rotation rate, $\Omega$, around the $\z$ axis, the following is an exact reformulation of the fully compressible inviscid momentum equations 
\Beq
\label{compressible-momentum-eq}
\pd{t}u + (\curl u + \f  \, \z) \times u + \grad \varpi \ = \ T\, \grad s. 
\Eeq
The variable $u$ is the fluid velocity, $T$ is temperature and $s$ is entropy per unit mass. 
\Eq{compressible-momentum-eq} replaces the pressure, $p$, with the Bernoulli function, $\varpi$,
\Beq
\frac{\dd{p}}{\rho} \ =  \ \dd{h} -    T \dd{s}, \qquad \varpi \ \equiv \ \frac{|u|^{2}}{2} + h + \phi,
\Eeq
where $h$ is the enthalpy per unit mass. 
Because the convection zone only contains 2\% of the solar mass, $M_{\sol}$  \cite{Gough2007}, the gravitational potential $\phi \approx - G M_{\sol}/r$, where $G$ is Newton's gravitational constant. 

A well-mixed convection zone implies a well-defined adiabatic stratification \cite{ChristensenDalsgaard2002}, which varies only with the gravitational potential,
\Beq
\dd{p_{0}} \ = \ - \rho_{0} \dd{\phi}  , \quad \dd{h_{0}} \ = \ - \dd{\phi} , \quad \dd{s_0} = 0. 
\label{background}
\Eeq
The exception happens in the upper $\O{2\%}$, where strong super-adiabatic stratification drives flow close to the speed of sound.

Assuming an ideal gas law, $h_{0} = c_{p} T_{0} = \gamma p_{0}/\rho_{0}/(\gamma-1)$.
The detailed specific heat depends on the ionisation fraction and elemental abundances, which only become significant effects very near the solar surface.
We assume constant values for  $c_{p}$ and $\gamma = 5/3$; see Table~\ref{table:parameters}.
Near the bottom of the convection zone $T_{0}(R_{\cz}) \approx 2.3\times 10^{6}\,$K.
Near the surface $T_{0}(R_{\sol})/T_{0}(R_{\cz}) \approx 10^{-3}$, which vanishes to a good approximation. 
Therefore 
\Beq
T_{0}(r) \ \approx \ \frac{\phi(R_{\sol}) - \phi(r)}{c_{p}} \ = \   T_{0}(R_{\cz})\frac{R_{\cz}}{H_{\cz}} \frac{R_{\sol}-r}{r}.
\Eeq
For an adiabatic reference state\footnote{All quantities with a `0'  subscript denote adiabatic background reference states.}, $\rho_{0}(r)  \propto T_{0}(r)^{3/2}$, $p_{0}(r) \propto T_{0}(r)^{5/2}$. 
The values at the base of the convection zone fix the constants of proportionality.

\subsection*{Momentum, mass and energy}

In the bulk of the convection zone, all thermodynamic variables fluctuate from their reference values by $\O{}10^{-6}$, which permits the anelastic approximation \cite{Gough1969,BrownEtal2012,VasilEtal2013}. 
Therefore, 
\Beq
\label{anelastic-momentum-eq}
\pd{t}u + (\curl u + \f \, \z) \times u + \grad \varpi \ = \  T_{0} \grad s .
\Eeq
The only difference between the fully compressible and anelastic momentum equation is the replacement $T \to T_{0}$. 
We omit an explicit treatment of magnetism in our analysis. 
At most we expect $|B| \lesssim \sqrt{\mu_{0} \rho_{0}} |u|$, which does not alter any of our conclusions.
The anelastic approximation also implies 
\Beq
\label{div-eq}
\div \left( \rho_{0} u \right) \ = \ 0.
\Eeq
Energy transport closes the system,
\Beq
\label{energy-eq}
\rho_{0}\, \pd{t}\!\left( T_{0} s  + |u|^{2}/2 \right) + \div \left( \rho_{0}  \varpi  u\right) \ = \  \div \left( K_{0} \grad T_{0} \right).
\Eeq
\Eq{energy-eq} omits the diffusion of thermal fluctuations the same way \eq{anelastic-momentum-eq} omits viscous dissipation. 
These effects are essential in a turbulent fluid, but only become important for microscopic scales.
We cannot, however, ignore the radiative diffusion of the background, which provides the heat that drives buoyancy \cite{LepotEtal2018}. 

\subsection*{Flux balance}

On average, energy transport requires 
\Beq
\label{transport}
\langle \varpi u_{r} \rangle \ = \ \frac{1}{\rho_{0}}\left[\frac{L_{\sol}}{4\pi r^{2} } + K_{0} \frac{d T_{0}}{dr}\right] \ \equiv \ F_{0}(r).
\Eeq
The angled bracket on the left-hand side of \eq{transport} represents an average over time and horizontal spherical surfaces. 
We model the conductivity with Kramers' opacity \cite{KpylEtal2019} law, 
\Beq
\label{opacity}
K_{0}(r)  \ = \ \frac{c_{p} L_{\sol}}{4\pi G M_{\sol}} \! \left[\frac{T_{0}(r)}{T_{0}(R_{\cz})}\right]^{7/2}\!,
\Eeq
which assumes $\rho_{0}^{2} \propto T_{0}^{3}$. 
The two terms within the square brackets in \eq{transport} partially cancel because $K_{0}\, d T_{0}/dr < 0$. 
By definition $F_{0}(R_{\cz})=0$. 
\Eq{transport} is the foundation for all estimates; the right-hand side contains only known parameters. 

The quantity $F_{0}(r)$ represents a radiative \textit{``flux debt''}. 
Below the convection zone, radiative diffusion carries the entire luminosity. 
At some point near $R_{\cz}$, the temperature gradient needed to carry the total flux becomes larger than the adiabatic gradient and the system becomes unstable. 
Convection transports the remainder of the heat after the background relaxes to an almost adiabatic state.
Notably, $F_{0}(r)$ is not constant in radius, which implies the following two comments:
\begin{itemize}
\setlength\itemsep{-0.1cm}
\item 
No matter what, convective time scales are much slower than rotational timescales \textit{within some distance of the radiative zone}. 
\item
The important question is how far into the convection zone can rotation dominate buoyancy?
\end{itemize}
\Eq{transport} carries dimensional units of velocity cubed. 
Standard mixing-length theory assumes $\u \sim F_{0}^{1/3}$ \cite{BohmVitense1958}. 
While sometimes $F_{0}^{1/3}$ gives the actual flow speed, more freedom exist in the rapidly rotating regime. 
We require a better understanding of $\varpi$.

\subsection*{Rossby numbers, length scales and Taylor-Proudman}

The Rossby number measures the non-dimensional ratio of convective-to-rotational acceleration. 
A common definition is the \textit{Bulk Rossby number} 
\Beq
\label{Ro-bulk}
\RoB \ \equiv \ \frac{\u}{\f  \, \h},
\Eeq
where $\u$ represents a typical velocity amplitude and $\h$ a density scale height, 
\Beq
\label{H-rho}
\h  \equiv  \frac{c^{2}}{g} \ \approx \ (\gamma-1) \frac{ (R_{\sol}-r) \, r}{R_{\sol}},
\Eeq
where $c^{2} = \gamma p_{0}/\rho_{0}$ and $g = G M_{\sol}/r^{2}$.
Many references alternatively use the convection zone depth $\H$ in place of the scale height \cite{FeatherstoneHindman2016b}, e.g, $\h(R_{\cz})/\H \approx 0.46$.
\Eq{Ro-bulk} represents the ratio of relative-to-background vorticity assuming convective flows vary on a length scale comparable to $\h$ (or $\H$). 
Flows undoubtedly fluctuate over entire domain, but $\RoB$ overestimates rotational influence if the actual energy-containing scale is smaller. 

Motivated by buoyancy considerations, the \textit{Convective Rossby number} is
\Beq
\label{Ro-conv}
\RoC \ \equiv \ \frac{\sqrt{T_{0} |\!\left< s-s_{0}\right>\! | }}{\f  \, \h},
\Eeq
where $\left<s-s_{0}\right>$ represents characteristic mean entropy differences. 
This definition is equivalent to that used in the Boussinesq setting and defined in terms of the Rayleigh, Taylor and Prandtl numbers, $\RoC \sim \sqrt{\mathrm{Ra}/\mathrm{Ta}/\mathrm{Pr}}$ \cite{JulienEtal1996}.

Assuming  $\L < \h $ represents the actual (yet unknown) dynamical scale, we define an alternative \textit{Dynamical Rossby number}
\Beq
\label{ro}
\RoD \ \equiv \ \frac{\u}{ \f \, \L } \ \approx \ \frac{| \curl u |}{\f }.
\Eeq
The dynamical Rossby number is the closest estimate of the actual ratio between local and background vorticity. 
All definitions relate to each other in some way and will coincide in the non-rotating regime.

The Taylor-Proudman theorem \cite{Proudman1916,Taylor1917} is an important constraint for rapidly rotating systems. 
The formal result comes from taking the curl of \eq{anelastic-momentum-eq} and neglecting all effects other than rotation,
\Beq
\label{simple-TPT}
\f\, \z \, \cdot \grad u  \ \approx \  0,
\Eeq
Strictly speaking, \eq{simple-TPT} is simply not true for convection. 
A generic system with boundaries cuts off the dynamics to fit within the domain; hence $| \z \, \cdot \grad u |   \sim  |  u |/ \H $. 
In a deep shell, $| \z \, \cdot \grad u |   \sim  |  u |/ \h $. 

The more subtle view of the Taylor-Proudman \textit{constraint} comes from considering the magnitude of the neglected terms. 
For $\RoD < 1$, there is substantial degree of \textit{anisotropy} between the $\z$ and perpendicular directions,
\Beq
\label{anisotropy}
\frac{\L}{\h}  \approx  \RoD \quad \iff \quad \L  \approx  \sqrt{ \frac{ \u \h}{\f} }.
\Eeq
\Eq{anisotropy} results from taking the curl of the momentum equations and balancing the resulting Coriolis terms against the nonlinear terms, \ie
\Beq
\f \left( \z\,\div u -  \z \cdot \grad  u \right)  \ \sim \   \curl [ (\curl u) \times u ],  
\label{curl equ} 
\Eeq
which is the CI part of the CIA balance.
The curl eliminates leading-order geostrophic terms.
\Eq{anisotropy} follows from estimating the left-hand side of \eq{curl equ} with $\f \u/\h$ and the right-hand side with $\u^{2}/\L^{2}$.

\Eq{anisotropy} gives a direct link between flow speed and scale in a rapidly rotating system. 
Determining if this relation holds requires considering convective energy and momentum balances in detail. 

This anisotropic picture of rotating convection has existed for many years in the Boussinesq convection community. 
The anisotropy principle is present in the linear-stability results in Chandrasekhar's famous book on hydrodynamic stability \cite{Chandra1961}. 
Stevenson 1979 \cite{Stevenson1979} expanded the understanding using linear theory.
Ingersoll and Pollard 1982 \cite{IngersollPollard1982} made a similar analysis applied to the interiors of planets. 
Sakai 1997 observed the phenomena clearly in laboratory experiments \cite{Sakai1997}. 
In a fully nonlinear and turbulent setting, Julien and coworkers have refined and promoted the principles of anisotropy and local Taylor-Proudman balance for strongly nonlinear flows since the mid-1990s \cite{JulienEtal1996,JulienEtal1998,SpragueEtal2006,JulienEtal2012,StellmachEtal2014,JulienEtal2016}; see Aurnou \etal 2020 \cite{AurnouEtal2020} for a current summary including experimental evidence. 
Detailed notions of rotational anisotropy have been slow to gain traction in the stratified stellar modelling community and astrophysics in general; although that has been changing in recent years \cite{BarkerEtal2014,IrelandBrowning2018,GuervillyEtal2019,AndersEtal2019,CurrieEtal2020,JermynEtal2020}.
Overall, it seems that a consensus is forming surrounding the proper mechanical and thermal balances in rapidly rotating convection. 

\subsection*{Pressure}

If we know the size of $\varpi$ in terms $\L$ and $\u$, then \eqs{transport}{anisotropy} give the characteristic magnitude of everything else. 
In fully compressible dynamics, the pressure (equivalently $\varpi$) is a genuinely dynamical variable; it requires its own initial condition. 
In low-Mach-number flows, the pressure becomes a Lagrange multiplier \cite{VasilEtal2013} that enforces the divergence-free condition by cancelling locally compressive terms. 
Finding an equation for $\varpi$ requires multiplying \eq{anelastic-momentum-eq} by $\rho_{0}$ and taking the divergence, 
which eliminates the time-evolution term and leaves an elliptic equation for $\varpi$. 
Because $\varpi$ appears linearly, we can decompose the total Bernoulli function as a sum of three ``partial pressures", 
\Beq
\varpi \ \equiv \ \varpi_{\mathrm{Kin.}} + \varpi_{\mathrm{Ther.}} + \varpi_{\mathrm{Rot.}}.
\Eeq
Each partial pressure responds to three separate sources
\Beq
\label{varpi-K}
\div \left ( \rho_{0} \grad \varpi_{\mathrm{Kin.}} \right) &=&  \div (\rho_{0}\, u \times (\curl u ) ), 
\\
\label{varpi-T}
\div  \left ( \rho_{0} \grad \varpi_{\mathrm{Ther.}} \right) &=& \div (\rho_{0}\,T_{0} \grad s  ),
\\
\label{varpi-R}
\div \left ( \rho_{0} \grad \varpi_{\mathrm{Rot.}} \right) &=&   \div (\f \, \rho_{0}\,   u \times \z  ).
\Eeq
Solving \eqss{varpi-K}{varpi-R} in practice also requires appropriate boundary conditions; each of the same magnitude as its corresponding bulk source.

We can estimate the size of each partial pressure from its source. 
First, 
\Beq
\label{varpi-KT-estimate}
| \varpi_{\mathrm{Kin.}} | \ \sim \ \u^{2}, \qquad | \varpi_{\mathrm{Ther.}} | \ \sim \ T_{0} |s |.
\Eeq
At this point, the magnitude of $|s|$ is unknown, but it cannot dominate all other terms.
If all three partial pressure are assumed roughly equal then $\u \sim F_{0}^{1/3}$.
This has been the traditional assumption in mixing-length theory. 
But it could also happen that the rotational (geostrophic) pressure dominates the kinetic pressure. 

\Eq{varpi-R} differs from \eqs{varpi-K}{varpi-T} in that the left- and right-hand sides contain different numbers of derivatives. 
Geostrophy selects a length scale,
\Beq
\label{varpi-R-estimate}
| \varpi_{\mathrm{Rot.}} | \ \sim \ \f \, \u \L \ \sim \ \sqrt{ \f \,\h \, \u^{3} }.
\Eeq
Altogether, we find two separate cases: 
CASE I in the slowly rotating scenario, and CASE II in the rapidly rotating scenario.
See Table~\ref{table:Rossby cases} for a summary. 

\subsection*{Entropy} 

The magnitude of buoyancy variation poses a subtle question.
The direct entropy evolution satisfies  
\Beq
\rho_{0} T_{0} \left(\pd{t} s  + u \cdot \grad s \right)  \ = \ \div ( K_{0} \grad T_{0}). \label{Entropy evolution}
\Eeq
The right-hand side radiative heating varies only in radius. 
The deviations from an adiabatic background therefore contain both mean and fluctuating parts; $s = \left<s\right>\!(r) + \tilde{s}(r,\theta,\varphi)$.

Convection happens because large-scale entropy gradients become unstable to growing fluctuations. 
The instability saturates from turbulent transport counteracting the advection of the background. 
Specifically, $u \cdot \grad \left<s\right> \sim u \cdot \grad \tilde{s}$, or
\Beq
\L\, \left<s-s_{0}\right>   \ \sim \   \h\ \tilde{s}. 
\Eeq
An estimate for $\tilde{s}$ follows from the curl of the momentum equations,
\Beq
\curl [ (\curl u) \times u ] \ \sim \ \grad T_{0} \times \grad \tilde{s},
\Eeq
which is the IA part of CIA balance. 
Therefore 
\Beq
T_{0} \, \tilde{s} \ \sim \ \frac{\u^{2}}{\RoD}, \qquad T_{0}  \left<s-s_{0}\right> \ \sim \ \frac{\u^{2}}{\RoD^{2}}.
\Eeq
Putting everything together, in the rapidly rotating regime \cite{AurnouEtal2020}, 
\Beq
\RoC \ \sim \ \frac{\RoB}{\RoD} \ \sim \ \RoD.
\Eeq
With all the theoretical elements in place, we apply our analysis to the solar convection zone.

\begin{table}[ht] 
\smallskip
\centering
\begin{tabular}{llll}
\trow{Quantity}{\mathrm{Symbol}}{Value}{Unit} 
\hline
\hline
\trow{ solar mass }      	  		{ M_{\sol} }            { \sn{2.0}{30} }    { kg } 		    \hline
\trow{ solar radius }      	 		{ R_{\sol} }            { \sn{6.9}{6} }     { m } 			\hline
\trow{ convection-zone radius } 	{ R_{\cz} }             { 0.7 }             { $R_{\sol}$ } 	\hline
\trow{ convection-zone depth }     	{ \H }                  { 0.3 }             { $R_{\sol}$ } 	\hline
\trow{ average rotation rate } 		{ \Omega } 		        { \sn{2.6}{-6} }	{ 1/s } 		\hline
\trow{ total luminosity }   		{ L_{\sol} }            { \sn{3.8}{26} }    { W } 		    \hline
\trow{ bottom temperature } 		{ T_{0}(R_{\cz})}      	{ \sn{2.3}{6} }		{ K } 		  	\hline
\trow{ bottom density } 			{ \rho_{0}(R_{\cz}) } 	{ 210 }			    { kg/m$^3$ } 	\hline
\trow{ bottom pressure } 			{ p_{0}(R_{\cz}) } 	    { \sn{6.7}{12} }	{ Pa } 		    \hline
\trow{ specific heat } 				{ c_{p} } 		    	{ \sn{3.4}{4} }		{ J/kg/K } 		\hline
\trow{ adiabatic exponent }      	{ \gamma }              { 5/3 }             {  } 			\hline
\hline  
\\
\end{tabular}
\caption{Approximate parameters used for estimation calculations.
The convection-zone parameters (c.z.) come from Gough 2007 \cite{Gough2007}; the remaining parameters come from standard values.}
\label{table:parameters} 
\end{table}

\begin{table}[ht]
\smallskip
\centering
\begin{tabular}{cccccc} 
CASE 
& $\RoD$                                
& $\u$ 
& $\L$ 
& $\t^{-1}$ 
& Condition 
\\ \hline \hline \\
I 
& $\frac{F_{0}^{1/3}}{\f \h}$ 
& $F_{0}^{1/3}$ 
& $\h$ 
& $\frac{F_{0}^{1/3}}{\h}$ 
& $\RoD > 1$ 
\\ \\ \hline \\ 
II   
& $\frac{F_{0}^{1/5}}{(\f \h)^{3/5}}$ 
& $(F_{0} \RoD)^{1/3}  $  
& $\h\, \RoD$ 
& $\frac{F_{0}^{1/3}}{\h \RoD^{2/3}}$ 
& $\RoD < 1$ 
\\ \\ \hline \hline \\
\end{tabular}
\caption{Possible dynamical scalings in the large- and small- Rossby number regimes.
$\RoD$ represents the Dynamical Rossby number; \ie the ratio of local vorticity to rotation rate.
$\u$, $\L$, $\t = \L/\u$ represent the respective flow amplitude, length scale and time scale.
}
\label{table:Rossby cases}
\end{table}

\section{Results}

Table~\ref{table:parameters} summarises the parameters used to estimate the Rossby number, convective velocities and length scales. 

\begin{figure}[ht]
\begin{center}
\includegraphics[scale=0.4]{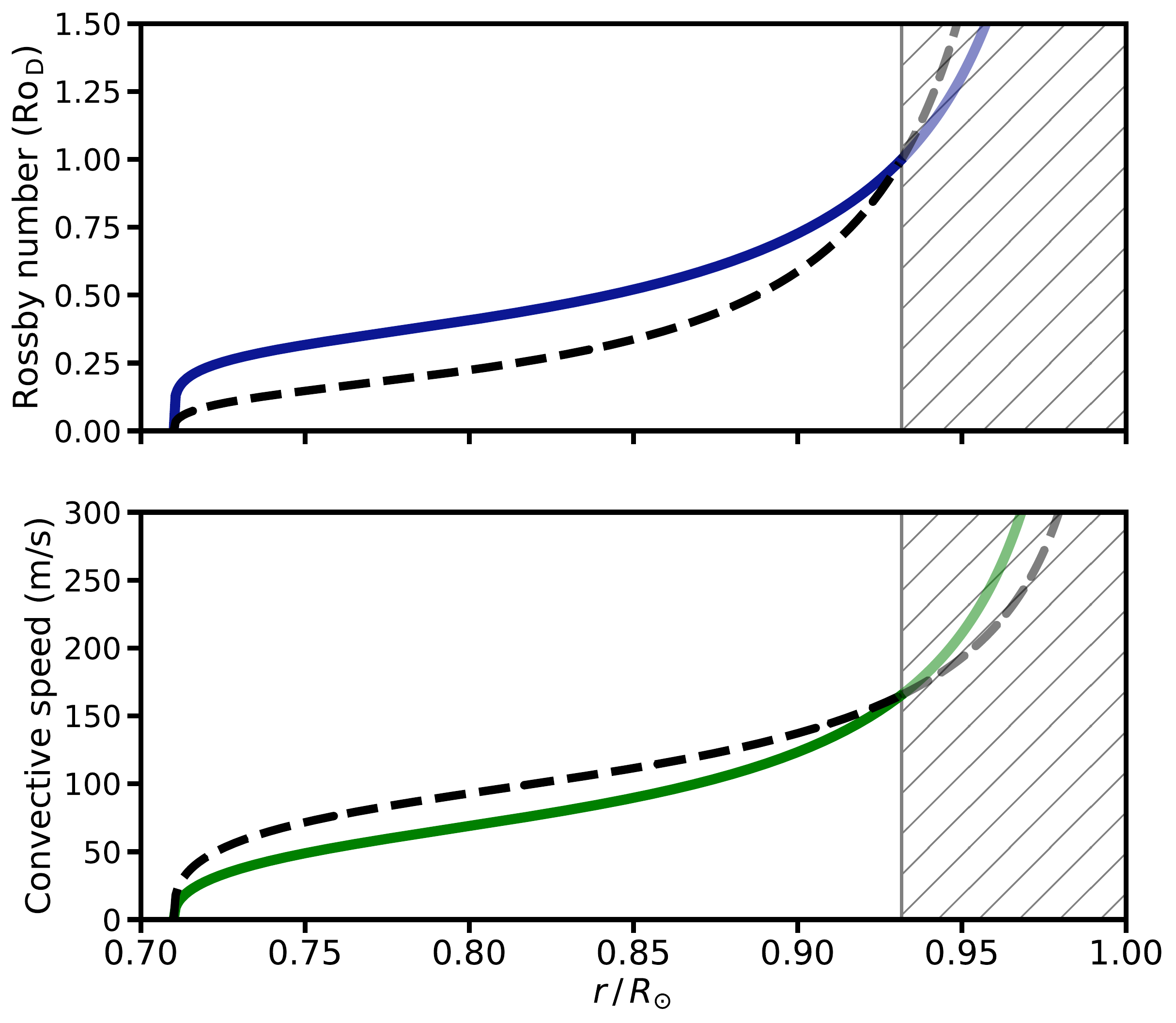}
\vspace{0.0 cm}
\caption{
Top: The estimated dynamical Rossby number as a function of depth. 
Bottom: The estimated convective flow speed estimates as a function of depth.
In both panels, the solid curves represent the rapidly rotating regime (CASE II).
The dashed lines show the slowly rotating counterfactual (CASE I).
Both $\RoD$ estimates are less than unity for much of the convection zone, hence rotating assumptions apply.
The shaded regions above $r/R_{\sol} \approx 0.93$ marks where rotational effects are subdominant. 
\label{fig:RoV}
}
\end{center}
\end{figure}

\Fig{fig:RoV}A shows Rossby number estimates as a function of depth. 
If we assume the traditional CASE I scenario, we find $\RoD < 1$ for $r/R_{\sol} \lesssim 0.93$. 
This is a contradiction because CASE I presumes no rotational influence, and hence CASE II must apply instead. 
While $\RoD$ is not asymptotically small, it is less the unity for most of the convection zone; and is less than 0.5 for the bottom half. 
\Fig{fig:RoV}B shows flow amplitude.
The flow speed is only slightly less in the rapidly rotating regime than the non-rotating equivalent. 
A weak power of the Rossby number distinguishes between CASE I \& II. 

\begin{figure}
\begin{center}
\includegraphics[scale=0.4]{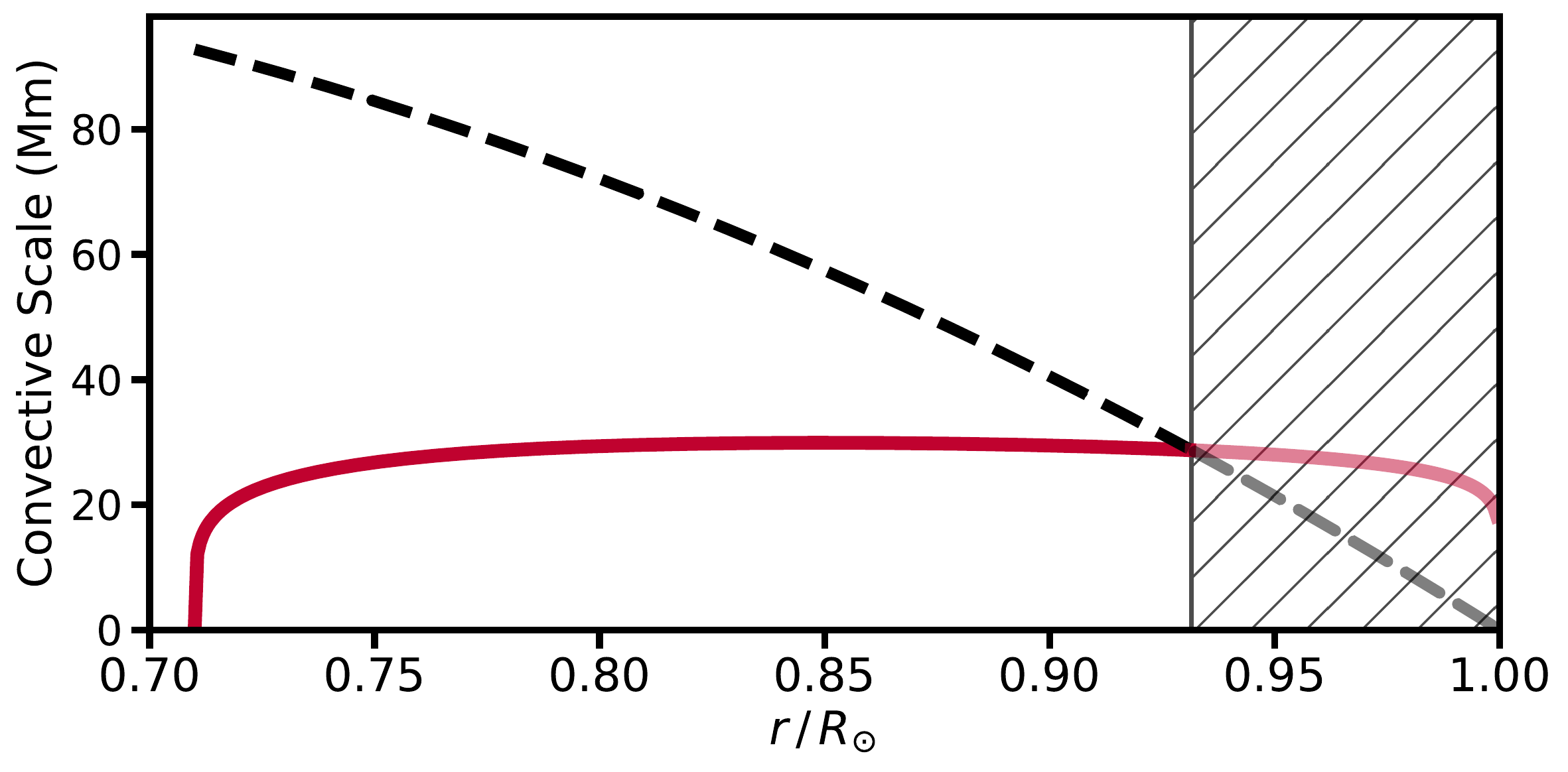}
\vspace{0.0 cm}
\caption{The convective length scale estimate as a function of depth in the solar convection zone.
The solid red curve shows the estimate under the rapidly rotating assumption (CASE II).
The dashed line shows the $\h(r)$ profile from \eq{H-rho} (CASE I).
The shaded region above $r/R_{\sol} \approx 0.93$ marks where rotation is subdominant; \ie a depth $\O{50\,Mm}$.
The rotationally constrained length scale stays consistently $\O{30 \,Mm}$ and is length scale realized below $r/R_{\sol} \approx 0.93$.
The density-scale height becomes the dominant length scale only in the near-surface regions.
\label{fig:length}}
\end{center}
\end{figure}

The dynamical length scale is the most dramatic effect of rapid rotation.
\Fig{fig:length} shows a consistent size for most of the convection zone, which is also much smaller than the bulk depth,
\Beq
\L \ \approx \ 0.15 \H \ \approx \ 30\, \mathrm{Mm}.
\Eeq
This number happens to coincide with the characteristic length scale of supergranulation; with spherical harmonic degree $\O{120-150}$.

\subsection*{Effects of geometry} 

A question remains how the misalignment of gravity and the rotation axis affects the estimates. 
Several studies investigated the dependence on the local latitude in Cartesian domains \cite{BrummellEtal1996,BrummellEtal1998,BrummellEtal2002,CurrieTobias2016,CurrieEtal2020}. 
No results seem definitive and much of the effort is still ongoing. 
Achieving rotational constraint at solar parameters requires a domain large enough to experience the variation of background forces. 
A sizable energy-containing eddy near the equator will feel off-equator effects. 

Table~\ref{table:Rossby cases}
depends on $H_{\rho}$, which entered the analysis by assuming $\hat{z} \cdot \grad \sim 1/\h \lesssim 1/\H$.
Variation along the rotation axis is the important overall reference length scale. 
The poles pose no challenge, where $\hat{r} \approx \hat{z}$. 
But starting from the equator, the distance from $r=R_{\cz}$ to $r=R_{\sol}$ in the $\hat{z}$ direction is
\Beq
H_{z} \ = \ \H \sqrt{ 2R_{\cz}/\H +1 } \  \approx  \ 5 \h.
\Eeq
We therefore expect at low latitudes modest increases to the true length scale appearing in Table~\ref{table:Rossby cases}, $\L \propto H_{z}^{2/5} \approx 1.9 \h^{2/5}$.
However $\RoD \propto H_{z}^{-3/5} \approx 0.4 \h^{-3/5}$
and $\u \propto H_{z}^{-1/5} \approx 0.7 \h^{-1/5}$.
Moreover, these estimates are the extreme case.
Therefore, near the equator, we might expect \Fig{fig:length} to show a slight increase in scale (perhaps up to $\O{50-60\,Mm}$ for $r \approx R_{\cz}$, but this would be accompanied by slower flows and additional rotational constraint.

\subsection*{Differential rotation}

Helioseismology provides a few additional checks on the overall rectitude of our estimates. 
Even with some differences, convection and shear provide proxies for one another \cite{JermynEtal2020}. 
Helioseismology, therefore, presents a consistency check for some of our estimates. 

As a function of radius, $r$, and co-latitude, $\theta$, the local rotation rate $\Omega(r,\theta)$ implies an angular inertial-frame bulk flow $u = r \sin \theta \, \Omega \, \hat{\varphi}$, which implies a total vorticity,
\Beq
\curl ( r \sin \theta \, \Omega \, \hat{\varphi} )  \ = \ \f\,\z  + r \! \sin \theta\, \grad \Omega \times \hat{\varphi}.
\Eeq
Assuming the differential rotation is strongly coupled to the dynamics, 
\Beq
\RoD \ \approx \ \frac{r \! \sin \theta  |\grad \Omega| }{2 \Omega }.
\Eeq

\begin{figure}[ht]
\begin{center}
\includegraphics[scale=1]{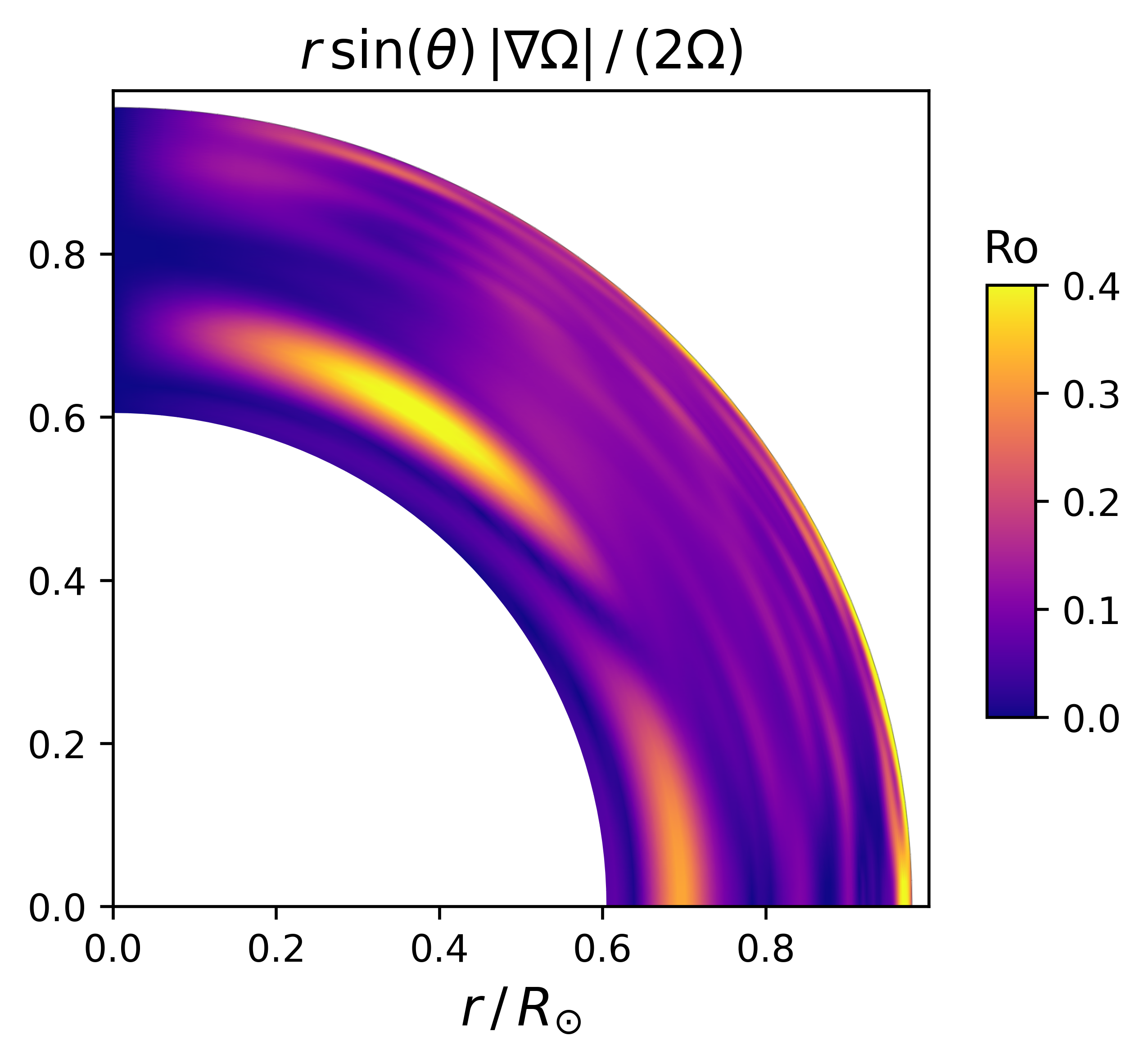}
\caption{
The local Rossby number computed from helioseismic-inferred profile $\Omega(r,\theta)$ in the convection zone. 
We use a 4th-order finite difference to compute the gradient from the raw rotation rate data; found in the electronic supplementary material from Larson and Schou 2018 \cite{LarsonSchou2018}.
\label{fig:2D-rossby}}
\end{center}
\end{figure}

\Fig{fig:2D-rossby} shows the local Rossby number of the differential rotation as a function of latitude and depth in the convection zone. 
The data comes from\footnote{Electronic Supplementary Material for the article ``Global-Mode Analysis of Full-Disk Data from the Michelson Doppler Imager and Helioseismic and Magnetic Imager.''} the local rotation rate in Larson and Schou 2018 \cite{LarsonSchou2018}.
We compute the gradient with a 4th-order finite difference derivative. 
A few pertinent observation are in order. 
The Rossby number is never more than 0.4 for $r/R_{\sol} < 0.95$. 
The tachocline ($0.65 < r/R_{\sol} <0.75$) dominates the picture in the deep interior. 
Above $r/R_{\sol} \approx 0.95$ the rapid increase of the Rossby number indicates the start of the near-surface shear layer, which is poorly understood. 
But for much of the bulk convection zone, the differential rotation Rossby number hovers consistently around $\O{0.1}$. 

Also intriguing, the half-width of the tachocline bump is $\O{35\,Mm}$, which is consistent with the $\O{30\,Mm}$ estimate for the convection. 
If convection maintains the tachocline, it seems reasonable that their sizes should match. 
We are aware that the resolution of helioseismology degrades with depth and tachocline widths are upper bounds on the actual thickness. 
Even so, the data from helioseismology accords with simple dynamical estimates.

\subsection*{Thermal wind}

The above picture of differential rotation is fully consistent with the large-scale thermal wind model of Balbus and coworkers (e.g., \cite{BalbusEtal2012}).
The above scalings allow a significant balance between, 
\Beq
r^{2} \sin(\theta) \, \pd{z} \Omega^{2}  \ \approx \ -T_{0}'(r)\, \pd{\theta} \tilde{s}. \label{Thermal-wind}
\Eeq
\Eq{Thermal-wind} does not necessarily give the exact large-scale entropy profile (e.g., due to possible Reynolds- and Maxwell-stress corrections), but it gives and good indication and provides a consistency check. 
It is also becoming clear how sensitive differential rotation can be to large-scale thermal gradients \cite{MatilskyEtal2020}.

\begin{figure}[ht]
\begin{center}
\includegraphics[scale=1]{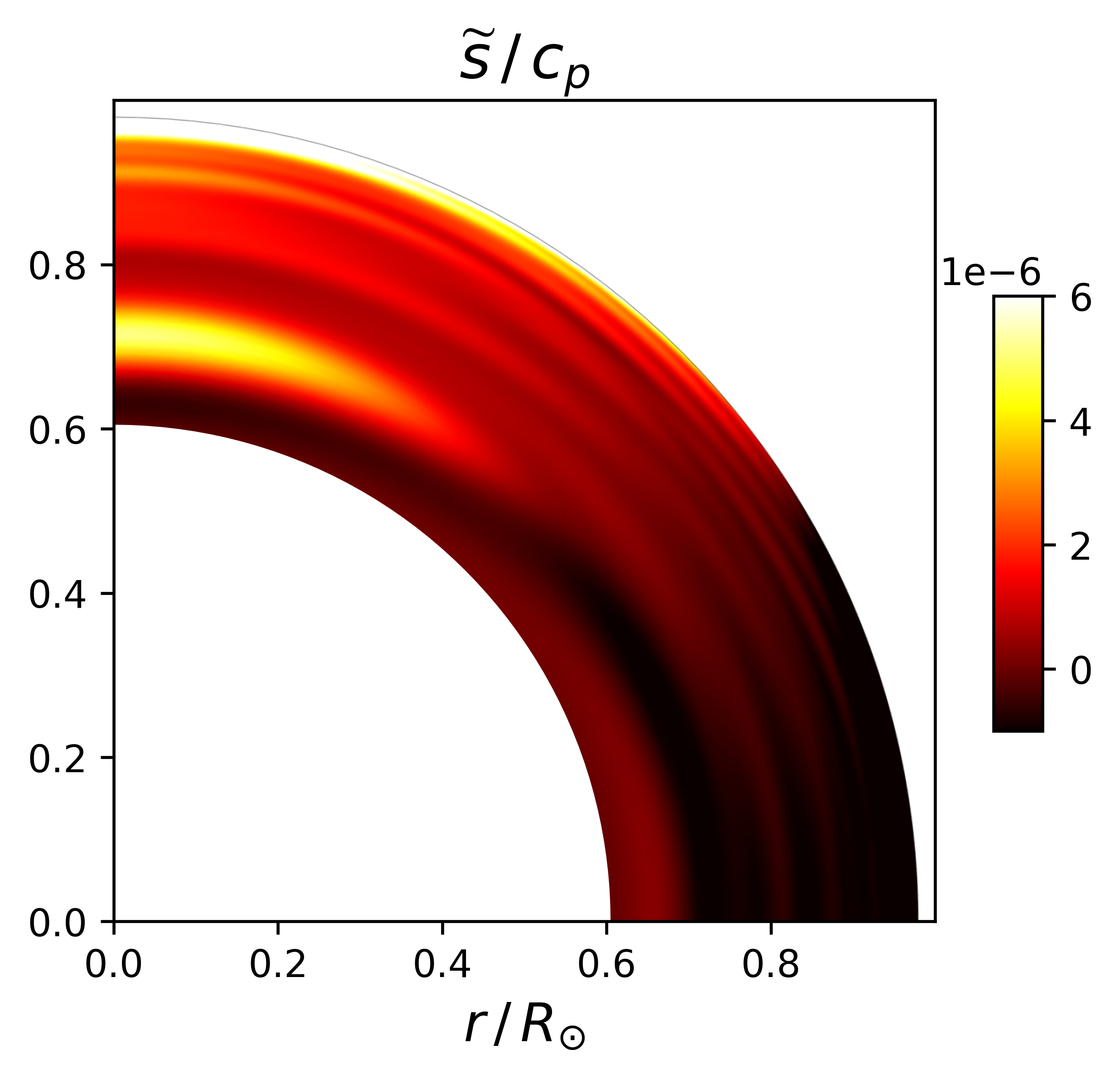}
\caption{
The local thermal-wind entropy profile computed from helioseismic-inferred profile $\Omega(r,\theta)$ in the convection zone. 
We use a 4th-order finite difference to compute $\pd{z}$ and a trapezoidal rule to integrate over $\theta$. The data is found in the electronic supplementary material from Larson and Schou 2018 \cite{LarsonSchou2018}.
\label{fig:2D-entropy}}
\end{center}
\end{figure}

\Fig{fig:2D-entropy} shows a solution to \eq{Thermal-wind}. 
We integrate the right-hand side over $\theta$ using the trapezoidal rule.  
As Balbus \etal 2012 \cite{BalbusEtal2012} pointed out, we can freely add any radial function to the solution.
We set the integration constant by $
\int_{0}^{\pi} \tilde{s}(r,\theta) \sin(\theta) \dd{\theta} = 0$.
The entropy state needed to maintain differential rotation is roughly the same needed to drive rotationally constrained turbulent convection.
Also intriguing, the entropy is better mixed near the equator than near the poles.

\section{Conclusions}

Based on well-understood physics, we furnish a detailed estimate of the degree of rotational constraint in the solar interior. 

We summarise our assumptions as follows
\begin{itemize}
\setlength\itemsep{-0.1cm}
\item Solar convection comprises a nearly adiabatic background hydrostatic balance, along with anelastic convective fluctuations and negligible viscous friction. 
\item Convective turbulence transports a radiative ``flux debt''; \ie the part of the solar luminosity that an adiabatic temperature gradient cannot carry.
\item The definition of ``rapid rotation'' is equivalent to leading-order balance between Coriolis and pressure forces (QG), followed by a joint triple balance between inertia, buoyancy and the non-divergent component of Coriolis deflections (CIA).
Magnetic energies should be roughly similar to kinetic energies.
\end{itemize}

We summarise our results as follows
\begin{itemize}
\setlength\itemsep{-0.1cm}
\item The solar convection zone is rotationally constrained roughly everywhere below the near-surface shear layer.
\item The flow amplitudes in the rotating regime are similar to what would exist in the absence of rotation. 
\item Rotation noticeably reduces the dynamical length scale.
We predict $\O{30\,Mm}$ robustly throughout the convection zone.
\item The dominant gradients should act perpendicular to the rotation axis, with scales $\sim \h  \O{100-200\,Mm}$ variation in the $\z$ direction. 
\item Individual deep convective structures should persist for multiple rotation periods.
\item Solar differential rotation is itself strongly rotationally influenced and indicative of the above conclusions. 
\item Both the equator and poles experience strong rotational influence.  
Moreover there should exist significant differences between the flow signatures observable in different regions. 
\end{itemize}

The $30\,\text{Mm}$ prediction is intriguing for several reasons.
First, the scale matches the observed size of surface supergranulation. 
While it is tempting to suggest supergranulation is the surface manifestation of deep convection, we do not believe the situation is altogether this simple.
While the spatial scales do match, the timescale of supergranulation is fast compared to rotation; \ie $\O{1\,day}$ versus $\O{1\,month}$. 
It also happens that the $\RoD \approx 1$ at $r/R_{\sol} \approx 0.93$, which is $\O{50\, Mm}$ below the surface.
While they may not be literally the same phenomena, a spatial matching means that the different phenomena can interact in interesting ways. 
We hope future studies will clarify the connections between anisotropic deep convection, supergranulation and the near-surface shear layer. 
All three processes surely interact in unforeseen ways with magnetism; e.g. \cite{Hotta2016,Hotta2018}.

We also believe our estimates resolve several past questions regarding observations and simulations. 
First, we believe the observation of Hanasoge \etal 2012, 2020 \cite{HanasogeEtal2012} find anomalously low signals because of a scale effect, rather than a genuinely small flow. 
The new results of Hanasoge \etal 2020 \cite{HanasogeEtal2020} also appear to imply this.
It is not clear what kind of signal should exist at $\O{200\,Mm}$ scales, given a true peak of power at $\O{30\,Mm}$ scales. 
But a direct and local inverse cascade should be weak. 
We anticipate at least an order-of-magnitude reduction.
Specifically, an inverse cascade should transfer energy to barotropic differential rotation; not to larger convective motions \cite{JulienEtal2018,CoustonEtal2020}. 

Our estimates also address numerous stellar-convection simulations carried out in recent decades. 
Any simulation with spherical harmonic degree less than $\O{1000}$ is unlikely to resolve a spectrum of low-Rossby, and large-Reynolds-number convection. 
It is not enough to only just resolve 30\,Mm structures. 
These features must also not dissipate viscously or thermally, as they surely do not in the real Sun.
We believe that the simulations of Featherstone and Hindman 2016 \cite{FeatherstoneHindman2016b} represent the first spherical simulations convincingly in this challenging computational regime.  
As computational tools improve, studies of this type are becoming increasingly more feasible (e.g., \cite{Hindman2020}).  
Moreover, it is possible that once the rotational length scales are safely captured in a low-friction regime, there may be diminishing returns from continued resolution increases. 

How can these predictions be tested against observations?
The expectation of vorticity parallel to the rotation axis offers a promising possibility.  
Unfortunately, helioseismic techniques are relatively insensitive to radial flow and are confined to regions within 60$^{\circ}$ of disk center.  
Helioseismically measuring vorticity parallel to the rotational-axis is quite difficult from the ecliptic plane.  
Techniques based on tracking supergranules fare a bit better away from disk center, and recent results appear to indicate the presence of large-scale vortical, cellular motions in the Sun's polar regions \cite{Hathaway2020}. 
Even better, observations from a polar vantage could use multiple techniques to analyse vorticity. 
Such high-latitude measurement are therefore becoming more crucial if theory and observations are to meet. 

Finally, we point out that the Sun rotated faster earlier in its history. 
It was also smaller and less luminous. 
All of these effects will surely produce an interesting interplay in the dynamical Rossby number, dominant length scales and hence magnetic field generation. 
However, global magnetism is the cause of rotational spindown.
The nature of magnetic breaking in stars is far from settled \cite{MetcalfeEtal2020}.
It all seems fascinatingly complicated. 
We hope our estimates motivate future studies in rotating and magnetised systems outside the present-day Sun.

\section*{Acknowledgements}

We thank Ben Brown for locating the differential rotation data used in this work, and also for helpful comments on manuscript. 
G.M.V. thanks Eliot Quataert for hosting in Berkeley in 2015 where much of this work was done, and also for helpful comments in the early stages. 
We also thank Brad Hindman for helpful conversations.

\footnotesize{

}


\begin{thebibliography}{99}

\bibitem{HanasogeEtal2012}
Shravan~M. Hanasoge, Thomas~L. Duvall, and Katepalli~R. Sreenivasan.
\newblock Anomalously weak solar convection.
\newblock {\em Proceedings of the National Academy of Sciences},
  109(30):11928--11932, June 2012.

\bibitem{GreerEtal2016}
Benjamin~J. Greer, Bradley~W. Hindman, and Juri Toomre.
\newblock Helioseismic imaging of supergranulation throughout the sun's
  near-surface shear layer.
\newblock {\em The Astrophysical Journal}, 824(2):128, June 2016.

\bibitem{HanasogeEtal2020}
Shravan~M. Hanasoge, Hideyuki Hotta, and Katepalli~R. Sreenivasan.
\newblock Turbulence in the sun is suppressed on large scales and confined to
  equatorial regions.
\newblock {\em Science Advances}, 6(30):eaba9639, July 2020.

\bibitem{MieschEtal2012}
Mark~S. Miesch, Nicholas~A. Featherstone, Matthias Rempel, and Regner
  Trampedach.
\newblock On the amplitude of convective velocities in the deep solar interior.
\newblock {\em The Astrophysical Journal}, 757(2):128, September 2012.

\bibitem{FeatherstoneHindman2016b}
Nicholas~A. Featherstone and Bradley~W. Hindman.
\newblock The emergence of solar supergranulation as a natural consequence of
  rotationally constrained interior convection.
\newblock {\em The Astrophysical Journal}, 830(1):L15, October 2016.

\bibitem{GilletJones2006}
N.~Gillet and C.~A. Jones.
\newblock The quasi-geostrophic model for rapidly rotating spherical convection
  outside the tangent cylinder.
\newblock {\em Journal of Fluid Mechanics}, 554(-1):343, April 2006.

\bibitem{SchwaigerEtal2019}
T~Schwaiger, T~Gastine, and J~Aubert.
\newblock Force balance in numerical geodynamo simulations: a systematic study.
\newblock {\em Geophysical Journal International},
  219(Supplement{\_}1):S101--S114, April 2019.

\bibitem{AurnouEtal2020}
Jonathan~M. Aurnou, Susanne Horn, and Keith Julien.
\newblock Connections between nonrotating, slowly rotating, and rapidly
  rotating turbulent convection transport scalings.
\newblock {\em Phys. Rev. Research}, 2:043115, Oct 2020.

\bibitem{Howe2009}
Rachel Howe.
\newblock Solar interior rotation and its variation.
\newblock {\em Living Reviews in Solar Physics}, 6, 2009.

\bibitem{ThompsonEtal2003}
Michael~J. Thompson, J{\o}rgen Christensen-Dalsgaard, Mark~S. Miesch, and Juri
  Toomre.
\newblock The internal rotation of the sun.
\newblock {\em Annual Review of Astronomy and Astrophysics}, 41(1):599--643,
  September 2003.

\bibitem{GastineEtal2013}
T.~Gastine, R.~K. Yadav, J.~Morin, A.~Reiners, and J.~Wicht.
\newblock From solar-like to antisolar differential rotation in cool stars.
\newblock {\em Monthly Notices of the Royal Astronomical Society: Letters},
  438(1):L76--L80, December 2013.

\bibitem{GastineEtal2014}
T.~Gastine, M.~Heimpel, and J.~Wicht.
\newblock Zonal flow scaling in rapidly-rotating compressible convection.
\newblock {\em Physics of the Earth and Planetary Interiors}, 232:36--50, July
  2014.

\bibitem{GuerreroEtal2013}
G.~Guerrero, P.~K. Smolarkiewicz, A.~G. Kosovichev, and N.~N. Mansour.
\newblock Differential rotation in solar-like stars from global simulations.
\newblock {\em The Astrophysical Journal}, 779(2):176, December 2013.

\bibitem{FeatherstoneMiesch2015}
Nicholas~A. Featherstone and Mark~S. Miesch.
\newblock Meridional circulation in solar and stellar convection zones.
\newblock {\em The Astrophysical Journal}, 804(1):67, May 2015.

\bibitem{HathawayEtal2003}
David~H. Hathaway, Dibyendu Nandy, Robert~M. Wilson, and Edwin~J. Reichmann.
\newblock Evidence that a deep meridional flow sets the sunspot cycle period.
\newblock {\em The Astrophysical Journal}, 589(1):665--670, may 2003.

\bibitem{HathawayEtal2010}
D.~H. Hathaway and L.~Rightmire.
\newblock Variations in the sun{'}s meridional flow over a solar
  cycle.
\newblock {\em Science}, 327(5971):1350--1352, March 2010.

\bibitem{Ulrich2010}
Roger~K. Ulrich.
\newblock Solar meridional circulation from {D}oppler shifts of the {F}e {I}
  line at 5250 {\aa} as measured by the 150-foot solar tower telescope at the
  {M}t.~{W}ilson observatory.
\newblock {\em The Astrophysical Journal}, 725(1):658--669, November 2010.

\bibitem{KholikovEtal2014}
S.~Kholikov, A.~Serebryanskiy, and J.~Jackiewicz.
\newblock Meridional flow in the solar convection zone. {I}. {M}easurements
  from {GONG} data.
\newblock {\em The Astrophysical Journal}, 784(2):145, March 2014.

\bibitem{JackiewiczEtal2015}
J.~Jackiewicz, A.~Serebryanskiy, and S.~Kholikov.
\newblock Meridional flow in the solar convection zone. {II.} {H}elioseismic
  inversions of {GONG} data.
\newblock {\em The Astrophysical Journal}, 805(2):133, May 2015.

\bibitem{GizonMC2020}
Laurent {Gizon}, Robert~H. {Cameron}, Majid {Pourabdian}, Zhi-Chao {Liang},
  Damien {Fournier}, Aaron~C. {Birch}, and Chris~S. {Hanson}.
\newblock {Meridional flow in the Sun{\textquoteright}s convection zone is a
  single cell in each hemisphere}.
\newblock {\em Science}, 368(6498):1469--1472, June 2020.

\bibitem{Ossendrijver2003}
Mathieu Ossendrijver.
\newblock The solar dynamo.
\newblock {\em Astronomy and Astrophysics Review}, 11(4):287--367, August 2003.

\bibitem{DikpatiMC99}
Mausumi {Dikpati} and Paul {Charbonneau}.
\newblock {A Babcock-Leighton Flux Transport Dynamo with Solar-like
  Differential Rotation}.
\newblock {\em The Astrophysical Journal}, 518(1):508--520, June 1999.

\bibitem{YeatesEtal2008}
Anthony~R. Yeates, Dibyendu Nandy, and Duncan~H. Mackay.
\newblock Exploring the physical basis of solar cycle predictions: Flux
  transport dynamics and persistence of memory in advection- versus
  diffusion-dominated solar convection zones.
\newblock {\em The Astrophysical Journal}, 673(1):544--556, January 2008.

\bibitem{MuozJaramilloEtal2009}
Andr{\'{e}}s Mu{\~{n}}oz-Jaramillo, Dibyendu Nandy, and Petrus C.~H. Martens.
\newblock Helioseismic data inclusion in solar dynamo models.
\newblock {\em The Astrophysical Journal}, 698(1):461--478, May 2009.

\bibitem{Moffatt1978}
K.~H. Moffatt.
\newblock {\em The Generation of Magnetic Fields in Electrically Conducting
  Fluids}.
\newblock Cambridge University Press, January 1978.

\bibitem{OlsonEtal1999}
Peter Olson, Ulrich Christensen, and Gary~A. Glatzmaier.
\newblock Numerical modeling of the geodynamo: Mechanisms of field generation
  and equilibration.
\newblock {\em Journal of Geophysical Research: Solid Earth},
  104(B5):10383--10404, May 1999.

\bibitem{Davidson2001}
P.~A. Davidson.
\newblock {\em An Introduction to Magnetohydrodynamics}.
\newblock Cambridge University Press, March 2001.

\bibitem{Hart1956}
A.~B. Hart.
\newblock Motions in the sun at the photospheric level: {VI}. large-scale
  motions in the equatorial region.
\newblock {\em Monthly Notices of the Royal Astronomical Society},
  116(1):38--55, February 1956.

\bibitem{LeightonEtal1962}
Robert~B. Leighton, Robert~W. Noyes, and George~W. Simon.
\newblock Velocity fields in the solar atmosphere. i. preliminary report.
\newblock {\em The Astrophysical Journal}, 135:474, March 1962.

\bibitem{HathawayEtal2000}
D.H. Hathaway, J.G. Beck, R.S. Bogart, K.T. Bachmann, G.~Khatri, J.M. Petitto,
  S.~Han, and J.~Raymond.
\newblock The photospheric convection spectrum.
\newblock {\em Solar Physics}, 193(1/2):299--312, 2000.

\bibitem{HathawayEtal2015}
David~H. Hathaway, Thibaud Teil, Aimee~A. Norton, and Irina Kitiashvili.
\newblock The sun's photospheric convection spectrum.
\newblock {\em The Astrophysical Journal}, 811(2):105, September 2015.

\bibitem{LawrenceEtal1999}
J.~K. Lawrence, A.~C. Cadavid, and A.~A. Ruzmaikin.
\newblock Characteristic scales of photospheric flows and their magnetic and
  temperature markers.
\newblock {\em The Astrophysical Journal}, 513(1):506--515, mar 1999.

\bibitem{AhlersEtal2009}
Guenter Ahlers, Siegfried Grossmann, and Detlef Lohse.
\newblock Heat transfer and large scale dynamics in turbulent
  rayleigh-b{\'{e}}nard convection.
\newblock {\em Reviews of Modern Physics}, 81(2):503--537, April 2009.

\bibitem{HathawayEtal2013}
D.~H. Hathaway, L.~Upton, and O.~Colegrove.
\newblock Giant convection cells found on the sun.
\newblock {\em Science}, 342(6163):1217--1219, December 2013.

\bibitem{Hathaway2020}
David~H. {Hathaway} and Lisa~A. {Upton}.
\newblock {Hydrodynamic Properties of the Sun's Giant Cellular Flows}.
\newblock {\em arXiv e-prints}, {}:arXiv:2006.06084, June 2020.

\bibitem{RinconEtal2017}
F.~Rincon, T.~Roudier, A.~A. Schekochihin, and M.~Rieutord.
\newblock Supergranulation and multiscale flows in the solar photosphere.
\newblock {\em Astronomy {\&} Astrophysics}, 599:A69, March 2017.

\bibitem{Lord14}
J.~W. {Lord}, R.~H. {Cameron}, M.~P. {Rast}, M.~{Rempel}, and T.~{Roudier}.
\newblock {The Role of Subsurface Flows in Solar Surface Convection: Modeling
  the Spectrum of Supergranular and Larger Scale Flows}.
\newblock {\em The Astrophysical Journal}, 793(1):24, September 2014.

\bibitem{DuvallEtal1993}
T.~L. Duvall, S.~M. Jeffferies, J.~W. Harvey, and M.~A. Pomerantz.
\newblock Time-distance helioseismology.
\newblock {\em Nature}, 362(6419):430--432, April 1993.

\bibitem{BasuEtal1999}
Sarbani Basu, H.~M. Antia, and S.~C. Tripathy.
\newblock Ring diagram analysis of near-surface flows in the sun.
\newblock {\em The Astrophysical Journal}, 512(1):458--470, feb 1999.

\bibitem{LindseyBraun1997}
C.~Lindsey and D.~C. Braun.
\newblock Helioseismic holography.
\newblock {\em The Astrophysical Journal}, 485(2):895--903, aug 1997.

\bibitem{BrunToomre2002}
Allan~Sacha Brun and Juri Toomre.
\newblock Turbulent convection under the influence of rotation: Sustaining a
  strong differential rotation.
\newblock {\em The Astrophysical Journal}, 570(2):865--885, may 2002.

\bibitem{MieschEtal2006}
Mark~S. Miesch, Allan~Sacha Brun, and Juri Toomre.
\newblock Solar differential rotation influenced by latitudinal entropy
  variations in the tachocline.
\newblock {\em The Astrophysical Journal}, 641(1):618--625, apr 2006.

\bibitem{Ghizaru10}
Mihai {Ghizaru}, Paul {Charbonneau}, and Piotr~K. {Smolarkiewicz}.
\newblock {Magnetic Cycles in Global Large-eddy Simulations of Solar
  Convection}.
\newblock {\em The Astrophysical Journal Letters}, 715(2):L133--L137, June
  2010.

\bibitem{Racine11}
{\'E}tienne {Racine}, Paul {Charbonneau}, Mihai {Ghizaru}, Am{\'e}lie
  {Bouchat}, and Piotr~K. {Smolarkiewicz}.
\newblock {On the Mode of Dynamo Action in a Global Large-eddy Simulation of
  Solar Convection}.
\newblock {\em The Astrophysical Journal}, 735(1):46, July 2011.

\bibitem{Brown11}
Benjamin~P. {Brown}, Mark~S. {Miesch}, Matthew~K. {Browning}, Allan~Sacha
  {Brun}, and Juri {Toomre}.
\newblock {Magnetic Cycles in a Convective Dynamo Simulation of a Young
  Solar-type Star}.
\newblock {\em The Astrophysical Journal}, 731(1):69, April 2011.

\bibitem{Kapyla12}
Petri~J. {K{\"a}pyl{\"a}}, Maarit~J. {Mantere}, and Axel {Brand enburg}.
\newblock {Cyclic Magnetic Activity due to Turbulent Convection in Spherical
  Wedge Geometry}.
\newblock {\em The Astrophysical Journal Letters}, 755(1):L22, August 2012.

\bibitem{Warnecke14}
J{\"o}rn {Warnecke}, Petri~J. {K{\"a}pyl{\"a}}, Maarit~J. {K{\"a}pyl{\"a}}, and
  Axel {Brandenburg}.
\newblock {On The Cause of Solar-like Equatorward Migration in Global
  Convective Dynamo Simulations}.
\newblock {\em The Astrophysical Journal Letters}, 796(1):L12, November 2014.

\bibitem{Nelson15}
Nicholas~J. {Nelson}, Benjamin~P. {Brown}, Allan~S. {Brun}, Mark~S. {Miesch},
  and Juri {Toomre}.
\newblock {Buoyant Magnetic Loops Generated by Global Convective Dynamo
  Action}.
\newblock {\em Solar Physics}, 289(2):441--458, February 2014.

\bibitem{Gastine13}
T.~{Gastine}, J.~{Wicht}, and J.~M. {Aurnou}.
\newblock {Zonal flow regimes in rotating anelastic spherical shells: An
  application to giant planets}.
\newblock {\em Icarus}, 225(1):156--172, July 2013.

\bibitem{Guerrero13}
G.~{Guerrero}, P.~K. {Smolarkiewicz}, A.~G. {Kosovichev}, and N.~N. {Mansour}.
\newblock {Differential Rotation in Solar-like Stars from Global Simulations}.
\newblock {\em The Astrophysical Journal}, 779(2):176, December 2013.

\bibitem{Kapyla14}
P.~J. {K{\"a}pyl{\"a}}, M.~J. {K{\"a}pyl{\"a}}, and A.~{Brand enburg}.
\newblock {Confirmation of bistable stellar differential rotation profiles}.
\newblock {\em Astronomy and Astrophysics}, 570:A43, October 2014.

\bibitem{BrunBrowning2017}
Allan~Sacha Brun and Matthew~K. Browning.
\newblock Magnetism, dynamo action and the solar-stellar connection.
\newblock {\em Living Reviews in Solar Physics}, 14(1), September 2017.

\bibitem{MetcalfeEtal2016}
Travis~S. Metcalfe, Ricky Egeland, and Jennifer van Saders.
\newblock Stellar evidence that the solar dynamo may be in transition.
\newblock {\em The Astrophysical Journal}, 826(1):L2, jul 2016.

\bibitem{Hartlep13}
T.~{Hartlep}, J.~{Zhao}, A.~G. {Kosovichev}, and N.~N. {Mansour}.
\newblock {Solar Wave-field Simulation for Testing Prospects of Helioseismic
  Measurements of Deep Meridional Flows}.
\newblock {\em The Astrophysical Journal}, 762(2):132, January 2013.

\bibitem{ZhaoEtal2013}
Junwei Zhao, R.~S. Bogart, A.~G. Kosovichev, T.~L. Duvall, and Thomas Hartlep.
\newblock Detection of equatorward meridional flow and evidence of double-cell
  meridional circulation inside the sun.
\newblock {\em The Astrophysical Journal}, 774(2):L29, August 2013.

\bibitem{Gough2007}
Douglas Gough.
\newblock An introduction to the solar tachocline.
\newblock In D.~W. Hughes, R.~Rosner, and N.~O. Weiss, editors, {\em The Solar
  Tachocline}, pages 3--30. Cambridge University Press, 2007.

\bibitem{ChristensenDalsgaard2002}
J{\o}rgen Christensen-Dalsgaard.
\newblock Helioseismology.
\newblock {\em Reviews of Modern Physics}, 74(4):1073--1129, November 2002.

\bibitem{Gough1969}
D.~O. Gough.
\newblock The anelastic approximation for thermal convection.
\newblock {\em Journal of the Atmospheric Sciences}, 26(3):448--456, May 1969.

\bibitem{BrownEtal2012}
Benjamin~P. Brown, Geoffrey~M. Vasil, and Ellen~G. Zweibel.
\newblock Energy conservation and gravity waves in sound-proof treatments of
  stellar interiors. part i. anelastic approximations.
\newblock {\em The Astrophysical Journal}, 756(2):109, August 2012.

\bibitem{VasilEtal2013}
Geoffrey~M. Vasil, Daniel Lecoanet, Benjamin~P. Brown, Toby~S. Wood, and
  Ellen~G. Zweibel.
\newblock Energy conservation and gravity waves in sound-proof treatments of
  stellar interiors. part ii. lagrangian-constrained analysis.
\newblock {\em The Astrophysical Journal}, 773(2):169, August 2013.

\bibitem{LepotEtal2018}
Simon Lepot, S{\'{e}}bastien Auma{\^{\i}}tre, and Basile Gallet.
\newblock Radiative heating achieves the ultimate regime of thermal convection.
\newblock {\em Proceedings of the National Academy of Sciences},
  115(36):8937--8941, August 2018.

\bibitem{KpylEtal2019}
P.~J. K\"{a}pyl\"{a}, M.~Viviani, M.~J. K\"{a}pyl\"{a}, A.~Brandenburg, and
  F.~Spada.
\newblock Effects of a subadiabatic layer on convection and dynamos in
  spherical wedge simulations.
\newblock {\em Geophysical {\&} Astrophysical Fluid Dynamics},
  113(1-2):149--183, February 2019.

\bibitem{BohmVitense1958}
E.~B\"{o}hm-Vitense.
\newblock Uber wasserstoffkonvektions-zone in sternen und leuchtkra ``fte''.
\newblock {\em Z. Astrophys}, 46:108--143, 1958.

\bibitem{JulienEtal1996}
K.~Julien, S.~Legg, J.~Mcwilliams, and J.~Werne.
\newblock Rapidly rotating turbulent rayleigh-b{\'{e}}nard convection.
\newblock {\em Journal of Fluid Mechanics}, 322:243--273, September 1996.

\bibitem{Proudman1916}
J.~Proudman.
\newblock On the motion of solids in a liquid possessing vorticity.
\newblock {\em Proceedings of the Royal Society of London. Series A, Containing
  Papers of a Mathematical and Physical Character}, 92(642):408--424, July
  1916.

\bibitem{Taylor1917}
G.~I. Taylor.
\newblock Motion of solids in fluids when the flow is not irrotational.
\newblock {\em Proceedings of the Royal Society of London. Series A, Containing
  Papers of a Mathematical and Physical Character}, 93(648):99--113, March
  1917.

\bibitem{Chandra1961}
S.~Chandrasekhar.
\newblock {\em Hydrodynamic and Hydromagnetic Stability}.
\newblock Oxford University Press, January 1961.

\bibitem{Stevenson1979}
David~J. Stevenson.
\newblock Turbulent thermal convection in the presence of rotation and a
  magnetic field: A heuristic theory.
\newblock {\em Geophysical {\&} Astrophysical Fluid Dynamics}, 12(1):139--169,
  January 1979.

\bibitem{IngersollPollard1982}
Andrew~P. Ingersoll and David Pollard.
\newblock Motion in the interiors and atmospheres of jupiter and saturn: scale
  analysis, anelastic equations, barotropic stability criterion.
\newblock {\em Icarus}, 52(1):62--80, October 1982.

\bibitem{Sakai1997}
Satoshi Sakai.
\newblock The horizontal scale of rotating convection in the geostrophic
  regime.
\newblock {\em Journal of Fluid Mechanics}, 333:85--95, February 1997.

\bibitem{JulienEtal1998}
Keith Julien, Edgar Knobloch, and Joseph Werne.
\newblock A new class of equations for rotationally constrained flows.
\newblock {\em Theoretical and Computational Fluid Dynamics}, 11(3-4):251--261,
  June 1998.

\bibitem{SpragueEtal2006}
Michael Sprague, Keith Julien, Edgar Knobloch, and Joseph Werne.
\newblock Numerical simulation of an asymptotically reduced system for
  rotationally constrained convection.
\newblock {\em Journal of Fluid Mechanics}, 551(-1):141, March 2006.

\bibitem{JulienEtal2012}
Keith Julien, Edgar Knobloch, Antonio~M. Rubio, and Geoffrey~M. Vasil.
\newblock Heat transport in low-rossby-number rayleigh-b{\'{e}}nard convection.
\newblock {\em Physical Review Letters}, 109(25), December 2012.

\bibitem{StellmachEtal2014}
S.~Stellmach, M.~Lischper, K.~Julien, G.~Vasil, J.{\hspace{0.167em}}S. Cheng,
  A.~Ribeiro, E.{\hspace{0.167em}}M. King, and J.{\hspace{0.167em}}M. Aurnou.
\newblock Approaching the asymptotic regime of rapidly rotating convection:
  Boundary layers versus interior dynamics.
\newblock {\em Physical Review Letters}, 113(25), December 2014.

\bibitem{JulienEtal2016}
Keith Julien, Jonathan~M. Aurnou, Michael~A. Calkins, Edgar Knobloch, Philippe
  Marti, Stephan Stellmach, and Geoffrey~M. Vasil.
\newblock A nonlinear model for rotationally constrained convection with ekman
  pumping.
\newblock {\em Journal of Fluid Mechanics}, 798:50--87, May 2016.

\bibitem{BarkerEtal2014}
Adrian~J. Barker, Adam~M. Dempsey, and Yoram Lithwick.
\newblock Theory and simulations of rotating convection.
\newblock {\em The Astrophysical Journal}, 791(1):13, July 2014.

\bibitem{IrelandBrowning2018}
Lewis~G. Ireland and Matthew~K. Browning.
\newblock The radius and entropy of a magnetized, rotating, fully convective
  star: Analysis with depth-dependent mixing length theories.
\newblock {\em The Astrophysical Journal}, 856(2):132, April 2018.

\bibitem{GuervillyEtal2019}
C{\'{e}}line Guervilly, Philippe Cardin, and Nathanaël Schaeffer.
\newblock Turbulent convective length scale in planetary cores.
\newblock {\em Nature}, 570(7761):368--371, June 2019.

\bibitem{AndersEtal2019}
Evan~H. Anders, Cathryn~M. Manduca, Benjamin~P. Brown, Jeffrey~S. Oishi, and
  Geoffrey~M. Vasil.
\newblock Predicting the rossby number in convective experiments.
\newblock {\em The Astrophysical Journal}, 872(2):138, feb 2019.

\bibitem{CurrieEtal2020}
Laura~K Currie, Adrian~J Barker, Yoram Lithwick, and Matthew~K Browning.
\newblock Convection with misaligned gravity and rotation: simulations and
  rotating mixing length theory.
\newblock {\em Monthly Notices of the Royal Astronomical Society},
  493(4):5233--5256, February 2020.

\bibitem{JermynEtal2020}
Adam~S Jermyn, Shashikumar~M Chitre, Pierre Lesaffre, and Christopher~A Tout.
\newblock Convective differential rotation in stars and planets {\textendash}
  i. theory.
\newblock {\em Monthly Notices of the Royal Astronomical Society},
  498(3):3758--3781, August 2020.

\bibitem{BrummellEtal1996}
Nicholas~H. Brummell, Neal~E. Hurlburt, and Juri Toomre.
\newblock Turbulent compressible convection with rotation. i. flow structure
  and evolution.
\newblock {\em The Astrophysical Journal}, 473(1):494--513, dec 1996.

\bibitem{BrummellEtal1998}
Nicholas~H. Brummell, Neal~E. Hurlburt, and Juri Toomre.
\newblock Turbulent compressible convection with rotation. {II}. mean flows and
  differential rotation.
\newblock {\em The Astrophysical Journal}, 493(2):955--969, feb 1998.

\bibitem{BrummellEtal2002}
Nicholas~H. Brummell, Thomas~L. Clune, and Juri Toomre.
\newblock Penetration and overshooting in turbulent compressible convection.
\newblock {\em The Astrophysical Journal}, 570(2):825--854, may 2002.

\bibitem{CurrieTobias2016}
Laura~K. Currie and Steven~M. Tobias.
\newblock Mean flow generation in rotating anelastic two-dimensional
  convection.
\newblock {\em Physics of Fluids}, 28(1):017101, January 2016.

\bibitem{LarsonSchou2018}
Timothy~P. Larson and Jesper Schou.
\newblock Global-mode analysis of full-disk data from the michelson doppler
  imager and the helioseismic and magnetic imager.
\newblock {\em Solar Physics}, 293(2), January 2018.

\bibitem{BalbusEtal2012}
Steven~A. Balbus, Henrik Latter, and Nigel Weiss.
\newblock Global model of differential rotation in the sun.
\newblock {\em Monthly Notices of the Royal Astronomical Society},
  420(3):2457--2466, January 2012.

\bibitem{MatilskyEtal2020}
Loren~I. Matilsky, Bradley~W. Hindman, and Juri Toomre.
\newblock Revisiting the sun's strong differential rotation along radial lines.
\newblock {\em The Astrophysical Journal}, 898(2):111, July 2020.

\bibitem{Hotta2016}
H.~{Hotta}, M.~{Rempel}, and T.~{Yokoyama}.
\newblock {Large-scale magnetic fields at high Reynolds numbers in
  magnetohydrodynamic simulations}.
\newblock {\em Science}, 351(6280):1427--1430, March 2016.

\bibitem{Hotta2018}
H.~{Hotta}.
\newblock {Breaking Taylor-Proudman Balance by Magnetic Fields in Stellar
  Convection Zones}.
\newblock {\em The Astrophysical Journall}, 860(2):L24, June 2018.

\bibitem{JulienEtal2018}
Keith Julien, Edgar Knobloch, and Meredith Plumley.
\newblock Impact of domain anisotropy on the inverse cascade in geostrophic
  turbulent convection.
\newblock {\em Journal of Fluid Mechanics}, 837, January 2018.

\bibitem{CoustonEtal2020}
Louis-Alexandre Couston, Daniel Lecoanet, Benjamin Favier, and Michael~Le Bars.
\newblock Shape and size of large-scale vortices: A generic fluid pattern in
  geophysical fluid dynamics.
\newblock {\em Physical Review Research}, 2(2), May 2020.

\bibitem{Hindman2020}
Bradley~W. {Hindman}, Nicholas~A. {Featherstone}, and Keith {Julien}.
\newblock Morphological classification of the convective regimes in rotating
  stars.
\newblock {\em The Astrophysical Journal}, 898(2):120, August 2020.

\bibitem{MetcalfeEtal2020}
Travis~S. Metcalfe, Jennifer~L. van Saders, Sarbani Basu, Derek Buzasi,
  William~J. Chaplin, Ricky Egeland, Rafael~A. Garcia, Patrick Gaulme, Daniel
  Huber, Timo Reinhold, Hannah Schunker, Keivan~G. Stassun, Thierry
  Appourchaux, Warrick~H. Ball, Timothy~R. Bedding, S{\'{e}}bastien Deheuvels,
  Luc{\'{\i}}a Gonz{\'{a}}lez-Cuesta, Rasmus Handberg, Antonio Jim{\'{e}}nez,
  Hans Kjeldsen, Tanda Li, Mikkel~N. Lund, Savita Mathur, Benoit Mosser,
  Martin~B. Nielsen, Anthony Noll, Zeynep~{\c{C}}elik Orhan, Sibel Örtel,
  {\^{A}}ngela R.~G. Santos, Mutlu Yildiz, Sallie Baliunas, and Willie Soon.
\newblock The evolution of rotation and magnetic activity in 94 aqr aa from
  asteroseismology with {TESS}.
\newblock {\em The Astrophysical Journal}, 900(2):154, sep 2020.

\end{thebibliography}
\end{document}